\documentclass[runningheads]{llncs}

\usepackage[T1]{fontenc}
\usepackage{graphicx}
\usepackage{tikz}
\usepackage{tabularx}
\usepackage{array,booktabs,multirow}
\usepackage{svg}
\usepackage{bbm}
\usepackage{amssymb}
\usepackage{cite}
\usepackage{mathrsfs}
\usepackage{amsmath}
\usepackage{graphicx}
\usepackage{graphics}
\usepackage{pgfplots}
\usepackage{bbm}
\usepackage{mathbbol}
\usepackage{enumitem}
\usepackage[normalem]{ulem}
\useunder{\uline}{\ul}{}
\setlist[itemize]{noitemsep, topsep=0pt, nosep, leftmargin=*}
\setlist[enumerate]{noitemsep, topsep=0pt, nosep, leftmargin=28pt}

\begin{document}
\title{SAFERec: Self-Attention and Frequency Enriched Model for Next Basket Recommendation}
%
%
%
%
%

\author{Oleg Lashinin\inst{1, 2} \and
Denis Krasilnikov\inst{2} \and
Aleksandr Milogradskii\inst{2} \and
Marina Ananyeva\inst{3,2}}

\authorrunning{O. Lashinin et al.}

\institute{Moscow Institute of Physics and Technology, Russia \and
T-Bank, Russia \and
National Research University Higher School of Economics, Russia
\email{fotol764@gmail.com}\\
\email{di.krasilnikov@gmail.com}\\
\email{alex.milogradsky@gmail.com}\\
\email{ananyeva.me@gmail.com}}

\maketitle              
\begin{abstract}
Transformer-based approaches such as BERT4Rec and SASRec demonstrate strong performance in Next Item Recommendation (NIR) tasks. However, applying these architectures to Next-Basket Recommendation (NBR) tasks, which often involve highly repetitive interactions, is challenging due to the vast number of possible item combinations in a basket. Moreover, frequency-based methods such as TIFU-KNN and UP-CF still demonstrate strong performance in NBR tasks, frequently outperforming deep-learning approaches. This paper introduces SAFERec, a novel algorithm for NBR that enhances transformer-based architectures from NIR by incorporating item frequency information, consequently improving their applicability to NBR tasks. Extensive experiments on multiple datasets show that SAFERec outperforms all other baselines, specifically achieving an 8\% improvement in Recall@10. 

\end{abstract}

\section{Introduction}

 \looseness -1 The field of next-basket recommendations (NBR) has attracted significant interest from both academic \cite{fpmc, li2023next, firstrepro} and industry \cite{kou2023modeling, katz2022learning}, primarily caused by the NBR's critical role in enhancing the user experience on e-commerce platforms. Predicting users' subsequent purchases is essential in today's digital shopping landscape, supporting features like item reminders \cite{lerche2016value}, product discovery \cite{jannach2021exploring}, and efficient order creation \cite{singh2020prediction, wang2022efficiently}. Additionally, the increasing complexity and size of e-commerce platforms demonstrate a strong demand for innovations and further improvements in NBR methodologies.

\looseness -1 Some of the best performing methods in NBR are frequency-based algorithms, such as TIFU-KNN\cite{tifuknn} and UP-CF \cite{upcf}, which outperform even deep learning methods on these tasks\cite{li2023next}. However, despite their superior performance, these methods have some limitations. For instance, they struggle to adapt to diverse data types, such as images or textual descriptions, and rely on user-based K-nearest Neighbors methods, leading to scalability issues in large datasets. 

\looseness -1 Furthermore, in recent years, transformer-based models \cite{vaswani2017attention} such as SASRec\cite{sasrec}, gSASRec\cite{gsasrec}, and BERT4Rec\cite{bert4rec} have demonstrated remarkable performance in next-item recommendations (NIR) owing to the ability to model the sequential nature of user's behavior and complex optimization objectives \cite{liu2024sequential, petrov2024aligning, sasrecrl}. However, applying these methods to NBR remains challenging because of the vast number of potential baskets. Nevertheless, the application of transformer-based algorithms remains a promising topic in NBR tasks, as advancements in these models could significantly improve the effectiveness of NBR methodologies.

In addition, recent studies \cite{abbattista2024enhancing, recanet, ren2019repeatnet} have highlighted the importance of integrating transformer-based and frequency-based methods. In particular, frequency-based methods can enhance user-specific recommendations by effectively capturing repeated items, which are crucial for NBR tasks. Our study shows that the item frequencies component is a necessary part that significantly boosts the performance of the standard SASRec model, at times even surpassing state-of-the-art methods on specific datasets. In this work, we propose a novel architecture for NBR called \textit{SAFERec} that combines transformer-based and frequency-based methods. To the best of our knowledge, there is a lack of work that utilizes frequency-aware components to adapt the NIR models for the NBR tasks.

\looseness -1 The contributions of this work are twofold. \textbf{First}, we propose a novel NBR method, SAFERec, which integrates a transformer layer and a frequency-aware module for NBR. To support reproducibility, we have publicly released the code of SAFERec\footnote{\url{https://github.com/anon-ecir-nbr/SAFERec}}. \textbf{Second}, we conduct offline experiments on three public datasets. The results show that SAFERec outperforms other state-of-the-art methods, specifically by up to 8\% in Recall@10, and recommends a more novel set of items.

\vspace*{-0.8em}
\section{SAFERec}

\textbf{Problem Formulation.} \, In a typical Next Basket Recommendation (NBR) scenario, we have a set of users $U$ and items $V$. For each user in the dataset there is an ordered set of purchases with baskets $B^{u} = {b_{1}^{u},\ldots, b_{|B^u|}^{u}}$, where $b_{k}^{u} = ({v^{u, j}_{1}, \ldots, v^{u, j}_{|b_{k}^{u}|} })$ represents an unordered set of items. The objective of NBR is to predict the next basket $b_{|B^{u}|+1}^{u}$ for a user $u$. In offline experiments, models usually predict $b_{|B^{u}|}^{u}$ based on the purchase history, known as the leave-one-out evaluation protocol. However, in real-world applications, the system operates with the entire $B^{u}$ and can be evaluated once the next basket $b_{|B^{u}|+1}^{u}$ is observed.

\subsection{Architecture} 
\textbf{History Encoding Module.} \, SAFERec processes each user's purchase history $B^{u}$ in mini-batches. This begins with the conversion of baskets into sparse multi-hot vectors $q_{k}^{u} \in \mathbb{R}^{|V|}$, where the $i$-th component equals 1 if item $i$ is present in basket $b_{k}^{u}$ and 0 otherwise. The user's entire purchase history $B^{u}$ is thus represented as a sparse matrix  $\mathbf{W_u} \in \mathbb{R}^{l_u \times |V|}$ of sparse vectors $q_{k}^{u}$, with $l_u$ denoting the number of purchases made by user $u$.

To handle variations in purchase history length, we introduce a hyperparameter $L$ representing the maximum number of recent baskets the model considers. If $L > l_{u}$, we apply left padding to align all user purchase histories to a uniform length. Conversely, if $L < l_{u}$, we select $L$ most recent purchase histories.

Given the typically large catalog size $|V|$, we utilize a series of fully connected (FC) layers, similar to the approach in Mult-VAE\cite{liang2018variational} with the $tanh$ activation function. For simplicity, the hyperparameter $d$ is employed for all hidden dimensions throughout these linear layers and other modules in the SAFERec. The final layer outputs the matrix $\mathbf{W_{u, lat}} \in \mathbb{R}^{L \times d}$, representing the user’s history, where each basket is encoded into a $d$-dimensional latent vector space.

\begin{figure}[t!]
\centering
\includegraphics[width=8cm]{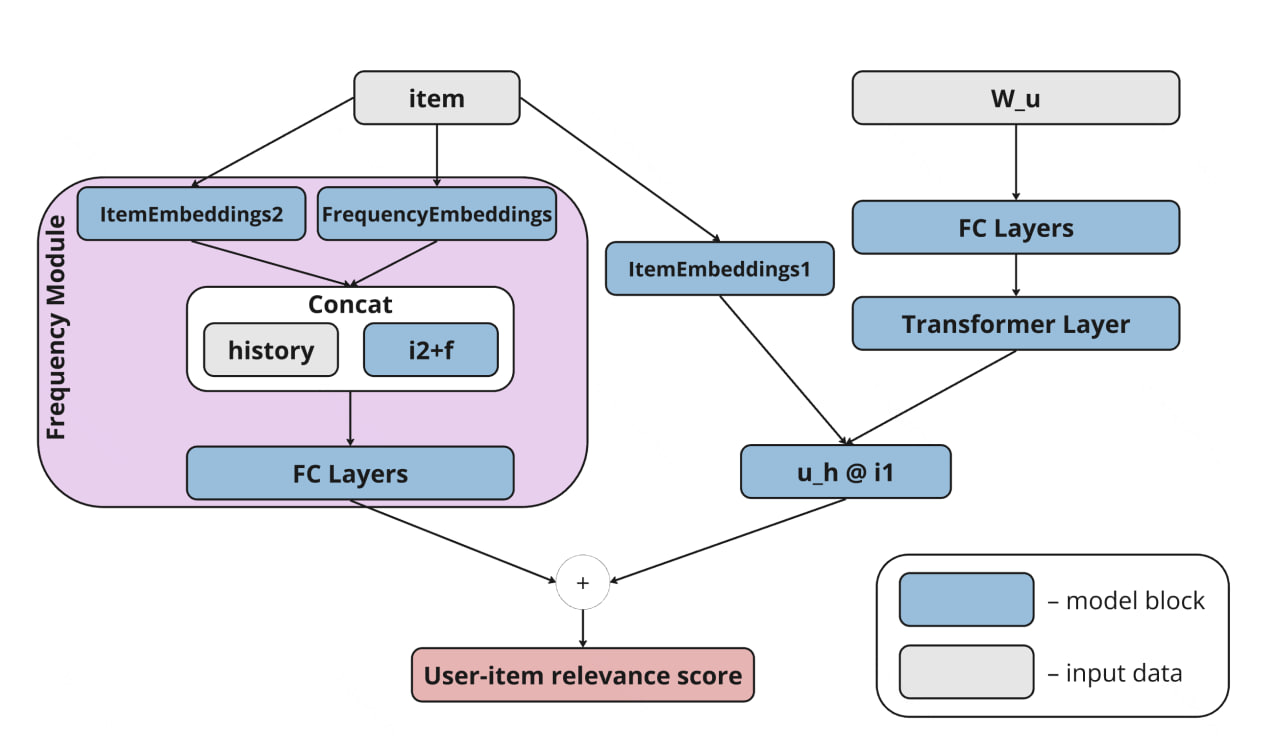}

\caption{SAFERec architecture overview.}

\label{fig:freq_overview}
\vspace{-2em}
\end{figure}

\looseness -1 \textbf{User Representation Module.} \, We employ a Transformer Layer \cite{vaswani2017attention} to capture user-specific representations, following a structure similar to SASRec \cite{sasrec} In contrast to the fixed positional embeddings used in \cite{vaswani2017attention}, our approach incorporates learned ones, denoted as $\mathbf{P} \in \mathbb{R}^{L \times d}$, for positions $1$ through $L$, resulting in $\mathbf{W_{u, p}} = \mathbf{W_{u, lat}} + \mathbf{P}$.  This change consistently delivered better results.

For hyperparameters, we adopt a setup similar to SASRec but vary the number of attention heads and stacked Transformer layers. Furthermore, we set inner hidden and output dimensions to $d$ to reduce the hyperparameter search space. The Transformer layer’s output, $\mathbf{W_{u, tr}} = Tr(\mathbf{W_{u, p}}) \in \mathbb{R}^{L \times d}$, encodes the user’s basket history $B^{u}$.
Following SASRec’s methodology, the final vector $w_u^{last}$ from $\mathbf{W_{u, tr}}$ acts as a hidden representation of the user. Finally, stacked FC layers are applied to $w_u^{last}$ to obtain the user representation $\mathbf{u_{h}}$. 

\looseness -1 \textbf{Approaches to Frequency Module.} \, To address the highly repetitive nature of user purchases in NBR, we introduce the Frequency Module (\textit{FM}). Prior work, such as TIFU-KNN\cite{tifuknn}, has shown that Recurrent Neural Networks struggle to accurately capture sums of input vectors, raising similar concerns about the effectiveness of Transformer layers for this task. In NBR, it is essential to recommend previously purchased items, yet encoding all possible item combinations within the latent user vector $\mathbf{u_{h}}$ can be challenging for models like SASRec.

\looseness -1 \textbf{Proposed FM Approach.} \, To tackle this issue, we incorporate item occurrence frequencies into the model, as illustrated in Figure \ref{fig:freq_overview}. SAFERec utilizes two learned item embedding matrices: $\mathbf{I_{1}}, \mathbf{I_{2}} \in \mathbb{R}^{|V| \times d}$, used for collaborative item representation and frequency aware representation, respectively. For each user-item pair, we define a history vector $\mathbf{h}_{u}^{i} = \left( \mathbb{1}_{\{i \in b_u^k\}} \right)_{k=1}^{L}$, indicating whether user $u$ purchased item $i$ in each of their past baskets $B^u$. This history vector helps capture the frequency of item interactions within the user's purchasing history.

\looseness -1 \textbf{Item-Specific and User-Specific Patterns.} \, \textit{FM} utilizes both user history vectors, $\mathbf{h}_{u}^{i}$, and item embedding, $i_2 \in \mathbf{I_{2}}$. For explicit frequency accounting, we use learned embeddings $FE \in R^{F_{\text{max}} \times d}$ for each frequency value from $0$ to $F_{max}$. All values greater than $F_{max}$ are clipped to $F_{\text{max}}$. We concatenate these components to form the vector $\mathbf{c}_{u}^{i} = \left[ i_2+f_i; \mathbf{h}_{u}^{i} \right]$, which we then process through a series of FC layers with a $\tanh$ activation function. This transformation outputs an item-specific frequency-based score $p_{ui}^i$ for each user-item pair.

\looseness -1  User preferences are incorporated similarly to SASRec through the dot product of the item embedding ${i}_1$ and the latent user vectors \(\mathbf{u_{h}}\) from the Transformer Layer, which encodes patterns from the user's interactions $\mathbf{h}_{u}^{i}$. The output $p_{uu}^i = \mathbf{i}_1 \cdot \mathbf{u_{h}}$ is the prediction score for a user-item pair, reflecting user-specific patterns.

\looseness -1 The final prediction of the SAFERec model is generated by summing both scores $p_u^i = p_{uu}^i + p_{ui}^i$. We argue that including frequency information helps the model generalize better, improving its performance compared to SASRec \cite{sasrec}.

\textbf{Objective Function.} \, In order to optimize the parameters of our neural network, we employ the Cross-Entropy Loss function, which remains a highly effective approach according to recent top-$n$ recommendation benchmarks \cite{sberpaper}. Like Mult-VAE \cite{liang2018variational}, which predicts the set of items in a top-n recommendation setup, SAFERec predicts the entire basket, in contrast to SASRec and other next-item models that typically predict only one item. Notably, incorporating item frequencies into variational inference methods is left for future work.

\section{Experiments}
We design our experiments to answer the following research questions:
\begin{enumerate}[font={\bfseries}, label={RQ\arabic*}]
\item How do frequency-aware techniques affect Transformer performance? 
\item How does SAFERec solve the NBR problem compared to well-established NBR baselines?
\end{enumerate}
\subsection{Offline Experimental setup}
\textbf{Datasets.} \, We run experiments on three public datasets: TaFeng\footnote{\url{https://www.kaggle.com/chiranjivdas09/ta-feng-grocery-dataset}}, Dunnhumby\footnote{\url{https://www.kaggle.com/datasets/frtgnn/dunnhumby-the-complete-journey}}, TaoBao\footnote{\url{https://tianchi.aliyun.com/dataset/649}}.
\looseness -1 We apply minimal preprocessing across all datasets: users and items with fewer than 5 interactions are removed, as well as users with only one basket. For TaoBao, this threshold is increased to 10 due to computational constraints. The main statistics are presented in Table \ref{tab:dataset}.

\textbf{Metrics.} \, We employ two ranking-based metrics: Recall@K and NDCG@K with binary relevance function. We also compute UserNovelty@K ($\text{UN}@K$), which indicates the rate of new items for user $u$, defined as $\text{UN}@K = \sum_{j=1}^{k} \mathbbm{1} [j \in B^{u}]$. All metrics are reported at cutoffs 10 and 100.

\looseness -1 \textbf{Baselines.} \, We compare SAFERec to well-established baselines for NBR. In a recent comprehensive NBR reproducibility study \cite{li2023next}, the authors show that TIFU-KNN \cite{tifuknn}, UP-CF \cite{upcf}, and DNNTSP \cite{dnntsp} - the latter being a deep learning method - are more stable compared to other proposed deep learning approaches. Therefore, we selected these models as our baselines.
Notably, we exclude novel methods such as ReCaNet \cite{recanet}, BTBR \cite{novelbasketrecsys23}, PerNIR \cite{ariannezhad2023personalizedPerNIR} as they address different issues such as repetition, next-novel, and within-basket recommendations. We omit the recent TIFU-KNN extension, TAIW \cite{romanov2023time}, which explicitly models time intervals. Although SAFERec could be extended to incorporate time-awareness, we leave this for future work. We also train a version of the SAFERec model without the FM, called SASRec* \cite{sasrec}, to represent its adaptation for the NBR task.

\textbf{Evaluation Protocol.} \, We adopt a leave-one-basket protocol, commonly used in recent studies on this task \cite{tifuknn, upcf}. The last basket for each user is assigned either for validation or testing, with half of the users allocated to the validation set and the other to the test set. All remaining baskets are used to train the models. For DNNTSP and SAFERec, we employ early stopping with a patience of 5 epochs and a maximum of 100 epochs. Hyperparameter tuning is conducted with Optuna\cite{optuna_2019}, using 50 trials for each model\footnote{The search grid and optimal hyperparameters are available in the repository: \url{https://github.com/anon-ecir-nbr/SAFERec/blob/main/README.md}}. This setup provides a robust foundation for evaluating SAFERec against the established baselines.

\begin{table}[t!]
\caption{Dataset statistics after preprocessing.}
\vspace{-1em}
\label{tab:dataset}
\begin{center}
\begin{tabular}{l|cccc}
\toprule
\multirow{2}{*}{Dataset} & \multirow{2}{*}{\#users} & \multirow{2}{*}{\#items} & \multirow{2}{*}{\#baskets} & \multirow{2}{*}{\begin{tabular}[c]{@{}c@{}}avg \#baskets\\ per user\end{tabular}} \\
                         &                          &                          &                            &                                                                                   \\ \hline
Dunnhumby                & 2495                     & 33910                    & 222196                     & 89.06                                                                             \\
Tafeng                   & 19112                    & 14713                    & 104257                     & 5.46                                                                              \\
TaoBao                   & 144408                   & 82310                    & 820764                     & 5.68                                                                              \\
\bottomrule 
\end{tabular}
\end{center}
\vspace{-2.5em}
\end{table}

\subsection{Results of Offline Experiments}
\begin{table}[t!]
\caption{The best and second-best models are bolded and underlined, respectively. The symbol $\dag$ denotes \textbf{no statistically significant difference} ($p > 0.05$) from the SAFERec model, as determined by a paired t-test with Bonferroni correction\cite{armstrong2014use}.}

\label{tab:results}
\resizebox{\textwidth}{!}{
\small
\begin{tabular}{c|l|ccccccc}
\toprule
\multirow{3}{*}{\rotatebox{90}{Dataset}}   & \multicolumn{1}{c|}{\multirow{3}{*}{Metric}} & \multicolumn{7}{c}{\multirow{2}{*}{Model}}                                                                                                                                                                                              \\
                           & \multicolumn{1}{c|}{}                        & \multicolumn{7}{c}{}                                                                                                                                                                                                                    \\
                           & \multicolumn{1}{c|}{}                        & \multicolumn{1}{c}{P-Pop} & \multicolumn{1}{c}{GP-Pop} & \multicolumn{1}{c}{TIFU-KNN} & \multicolumn{1}{c}{UP-CF} & \multicolumn{1}{c}{DNNTSP} & \multicolumn{1}{c}{SASRec*} & \multicolumn{1}{c}{SAFERec}                               \\ \hline
\multirow{6}{*}{\rotatebox{90}{TaFeng}}    & Recall@10                                    & 0.0633                    & 0.09991                    & 0.0897                       & 0.1214                    & \textbf{0.1276}$\dag$            & 0.0711                     & \underline{0.1256} ($\bigtriangledown$ 1\%) \\
                           & Recall@100                                   & 0.1845                    & 0.2708                     & 0.1557                       & {\ul 0.2701}              & 0.2455                     & 0.1947                     & \textbf{0.2761} ($\triangle$ 3\%)        \\
                           & NDCG@10                                      & 0.1589                    & 0.2084                     & 0.2005                       & {\ul 0.2417}              & 0.1955                     & 0.1381                     & \textbf{0.2504} ($\triangle$ 3\%)        \\
                           & NDCG@100                                     & 0.2143                    & 0.2642                     & 0.2254                       & {\ul 0.2854}              & 0.2524                     & 0.0995                     & \textbf{0.2927} ($\triangle$ 4\%)        \\
                           & UN@10                                        & 0.9136                    & 0.1373                     & 0.1373                       & 0.1805                    & {\ul 0.191}                & 0.9142                     & \textbf{0.2865} ($\triangle$ 50\%)       \\
                           & UN@100                                       & 0.9642                    & 0.7518                     & 0.7518                       & 0.7522                    & {\ul 0.7576}               & 0.9646                     & \textbf{0.7964} ($\triangle$ 5\%)        \\ \hline
\multirow{6}{*}{\rotatebox{90}{TaoBao}}    & Recall@10                                    & 0.0033                    & 0.0637                     & \textbf{0.0721}$\dag$                       & 0.0648                    &  0.0623               & 0.0046                     & \underline{0.0707}  ($\bigtriangledown$ 2\%)                                               \\
                           & Recall@100                                   & 0.0229                     & \textbf{0.1028}$\dag$            & 0.085                        & 0.1018                    & 0.0966                     & 0.0225                     & {\ul 0.1014}  ($\bigtriangledown$ 1\%)                                                 \\
                           & NDCG@10                                      & 0.0052                    & 0.067                      & \textbf{0.0843}              & 0.0684                    & 0.0639                     & 0.0065                     & {\ul 0.08} ($\bigtriangledown$ 4\%)                                                 \\
                           & NDCG@100                                     & 0.0157                    & 0.0835                     & 0.0896                       & {\ul 0.0922}$\dag$              & 0.0787                     & 0.0159                     & \textbf{0.0949} ($\triangle$ 1\%)                                           \\
                           & UN@10                                        & 0.9934                    & 0.0701                     & 0.0701                       & 0.0704                    & {\ul 0.0711}               & 0.9936                     & \textbf{0.0794}   ($\triangle$ 12\%)                                            \\
                           & UN@100                                       & 0.9964                    & 0.8627                     & 0.8627                       & 0.8627                    & {\ul 0.8629}               & 0.9965                     & \textbf{0.9406} ($\triangle 10\%$)                                              \\ \hline
\multirow{6}{*}{\rotatebox{90}{Dunnhumby}} & Recall@10                                    & 0.0814                    & 0.1415                     & 0.1472                       & {\ul 0.1503}              & 0.1279                     & 0.0840                     & \textbf{0.1619}  ($\triangle$ 8\%)                                              \\
                           & Recall@100                                   & 0.1815                    & 0.3312                     & 0.3402                       & {\ul 0.3395}              & 0.3023                     & 0.1921                     & \textbf{0.3585}  ($\triangle$ 6\%)                                              \\
                           & NDCG@10                                      & 0.2703                    & 0.3805                     & 0.3913                       & {\ul 0.4026}$\dag$              & 0.3492                     & 0.2655                     & \textbf{0.4086}    ($\triangle$ 1\%)                                           \\
                           & NDCG@100                                     & 0.2812                    & 0.394                      & 0.4035                       & {\ul 0.4063}$\dag$              & 0.3961                     & 0.2857                     & \textbf{0.4184}   ($\triangle$ 2\%)                                            \\
                           & UN@10                                        & 0.4771                    & 0.0033                     & 0.0033                       & 0.0033                    & {\ul 0.0108}               & 0.4507                     & \textbf{0.0147}  ($\triangle$ 36\%)                                             \\
                           & UN@100                                       & 0.6707                    & 0.0555                     & 0.0555                       & 0.0764                    & {\ul 0.0923}               & 0.6659                     & \textbf{0.1029}     ($\triangle$ 11\%)                                          \\
\bottomrule
\end{tabular}
}
\vspace{-1.7em}
\end{table}

Table \ref{tab:results} presents findings for both \textbf{RQ1} and \textbf{RQ2}. First, SAFERec consistently outperforms the SASRec model \cite{sasrec}, demonstrating its ability to capture item-specific patterns of recurrent behavior effectively. Additionally, SASRec*, performs nearly at the P-Pop level, suggesting it struggles to promote previously consumed items through hidden vectors alone. Although SASRec* shows an advantage in $\text{UN}@K$, this metric is less emphasized due to its poor ranking quality. Nevertheless, SAFERec excels in recommending novel items, offering a more diverse user experience with higher accuracy. This balance means users receive accurate recommendations while also discovering new products.

\looseness -1 Furthermore, Table \ref{tab:results} addresses \textbf{RQ2}, showing that SAFERec outperforms baselines across all datasets, except the TaoBao dataset, where GP-Pop and TIFU-KNN surpass the model in Recall@100 and NDCG@10, likely due to the dataset's emphasis on rare items. GP-Pop and TIFU-KNN recommend rare items based on user purchase history rather than learned hidden representations. Notably, baseline results align with reproducibility papers \cite{liu2024measuring, li2023next}, where TIFU-KNN, UP-CF, and DNNTSP show comparable performance and outperform GP-Pop. Despite accuracy differences, $\text{UN}@K$ indicates SAFERec's higher rate of recommending new items, which could enhance user satisfaction and engagement, outlined in \cite{herlocker2004evaluating, kaminskas2016diversity, novelbasketrecsys23}. Additionally, SAFERec maintains competitive accuracy metrics while exploring novel items. For instance, on TaFeng, SAFERec achieves better quality metrics than UP-CF, recommending 28\% more novel items in the top 10 compared to UP-CF's 18\%. These results show SAFERec's ability to take exploratory risks without compromising ranking quality, a trend consistent across datasets.


\section{Conclusions}

Our study introduces SAFERec, a novel model for the next-basket recommendations. Inspired by SASRec, designed for the next-item prediction, we present a transformer-based architecture incorporating user frequencies of purchasing particular items. Our experiments on three open-source datasets demonstrate that SAFERec outperforms state-of-the-art methods for the next-basket recommendations, achieving improvements of up to 8\% in Recall@10 and up to 50\% in the $\text{UN}@10$ metric. Additionally, SAFERec surpasses TIFU-KNN by as much as 56\% in Recall@100, all while maintaining scalability in large datasets through an efficient mini-batch training approach. We hope our SAFERec model will encourage researchers and practitioners to explore the potential of transformer-based methods in next-basket recommendations.

%
%

\bibliographystyle{splncs04}
\bibliography{sample-ceur}

\begin{thebibliography}{10}
\providecommand{\url}[1]{\texttt{#1}}
\providecommand{\urlprefix}{URL }
\providecommand{\doi}[1]{https://doi.org/#1}

\bibitem{abbattista2024enhancing}
Abbattista, D., Anelli, V.W., Di~Noia, T., Macdonald, C., Petrov, A.V.: Enhancing sequential music recommendation with personalized popularity awareness. In: Proceedings of the 18th ACM Conference on Recommender Systems. p. 1168–1173. RecSys '24, Association for Computing Machinery, New York, NY, USA (2024). \doi{10.1145/3640457.3691719}, \url{https://doi.org/10.1145/3640457.3691719}

\bibitem{optuna_2019}
Akiba, T., Sano, S., Yanase, T., Ohta, T., Koyama, M.: Optuna: A next-generation hyperparameter optimization framework. In: Proceedings of the 25th {ACM} {SIGKDD} International Conference on Knowledge Discovery and Data Mining (2019)

\bibitem{recanet}
Ariannezhad, M., Jullien, S., Li, M., Fang, M., Schelter, S., de~Rijke, M.: Recanet: A repeat consumption-aware neural network for next basket recommendation in grocery shopping. In: Proceedings of the 45th International ACM SIGIR Conference on Research and Development in Information Retrieval. pp. 1240--1250 (2022)

\bibitem{ariannezhad2023personalizedPerNIR}
Ariannezhad, M., Li, M., Schelter, S., De~Rijke, M.: A personalized neighborhood-based model for within-basket recommendation in grocery shopping. In: Proceedings of the Sixteenth ACM International Conference on Web Search and Data Mining. pp. 87--95 (2023)

\bibitem{armstrong2014use}
Armstrong, R.A.: When to use the b onferroni correction. Ophthalmic and Physiological Optics  \textbf{34}(5),  502--508 (2014)

\bibitem{upcf}
Faggioli, G., Polato, M., Aiolli, F.: Recency aware collaborative filtering for next basket recommendation. In: Proceedings of the 28th ACM Conference on User Modeling, Adaptation and Personalization. pp. 80--87 (2020)

\bibitem{herlocker2004evaluating}
Herlocker, J.L., Konstan, J.A., Terveen, L.G., Riedl, J.T.: Evaluating collaborative filtering recommender systems. ACM Transactions on Information Systems (TOIS)  \textbf{22}(1),  5--53 (2004)

\bibitem{tifuknn}
Hu, H., He, X., Gao, J., Zhang, Z.L.: Modeling personalized item frequency information for next-basket recommendation. In: Proceedings of the 43rd international ACM SIGIR conference on research and development in information retrieval. pp. 1071--1080 (2020)

\bibitem{jannach2021exploring}
Jannach, D., Jesse, M., Jugovac, M., Trattner, C.: Exploring multi-list user interfaces for similar-item recommendations. In: Proceedings of the 29th ACM Conference on User Modeling, Adaptation and Personalization. pp. 224--228 (2021)

\bibitem{kaminskas2016diversity}
Kaminskas, M., Bridge, D.: Diversity, serendipity, novelty, and coverage: a survey and empirical analysis of beyond-accuracy objectives in recommender systems. ACM Transactions on Interactive Intelligent Systems (TiiS)  \textbf{7}(1),  1--42 (2016)

\bibitem{sasrec}
Kang, W.C., McAuley, J.: Self-attentive sequential recommendation. In: 2018 IEEE international conference on data mining (ICDM). pp. 197--206. IEEE (2018)

\bibitem{katz2022learning}
Katz, O., Barkan, O., Koenigstein, N., Zabari, N.: Learning to ride a buy-cycle: A hyper-convolutional model for next basket repurchase recommendation. In: Proceedings of the 16th ACM Conference on Recommender Systems. pp. 316--326 (2022)

\bibitem{kou2023modeling}
Kou, Z., Manchanda, S., Lin, S.T., Xie, M., Wang, H., Zhang, X.: Modeling sequential collaborative user behaviors for seller-aware next basket recommendation. In: Proceedings of the 32nd ACM International Conference on Information and Knowledge Management. pp. 1097--1106 (2023)

\bibitem{lerche2016value}
Lerche, L., Jannach, D., Ludewig, M.: On the value of reminders within e-commerce recommendations. In: Proceedings of the 2016 Conference on User Modeling Adaptation and Personalization. pp. 27--35 (2016)

\bibitem{novelbasketrecsys23}
Li, M., Ariannezhad, M., Yates, A., de~Rijke, M.: Masked and swapped sequence modeling for next novel basket recommendation in grocery shopping. In: Proceedings of the 17th ACM Conference on Recommender Systems. pp. 35--46 (2023)

\bibitem{li2023next}
Li, M., Jullien, S., Ariannezhad, M., de~Rijke, M.: A next basket recommendation reality check. ACM Transactions on Information Systems  \textbf{41}(4),  1--29 (2023)

\bibitem{liang2018variational}
Liang, D., Krishnan, R.G., Hoffman, M.D., Jebara, T.: Variational autoencoders for collaborative filtering. In: Proceedings of the 2018 world wide web conference. pp. 689--698 (2018)

\bibitem{liu2024measuring}
Liu, Y., Li, M., Ariannezhad, M., Mansoury, M., Aliannejadi, M., de~Rijke, M.: Measuring item fairness in next basket recommendation: A reproducibility study. In: European Conference on Information Retrieval. pp. 210--225. Springer (2024)

\bibitem{liu2024sequential}
Liu, Z., Liu, S., Zhang, Z., Cai, Q., Zhao, X., Zhao, K., Hu, L., Jiang, P., Gai, K.: Sequential recommendation for optimizing both immediate feedback and long-term retention. arXiv preprint arXiv:2404.03637  (2024)

\bibitem{petrov2024aligning}
Petrov, A., Macdonald, C.: Aligning gptrec with beyond-accuracy goals with reinforcement learning. arXiv preprint arXiv:2403.04875  (2024)

\bibitem{gsasrec}
Petrov, A.V., Macdonald, C.: gsasrec: Reducing overconfidence in sequential recommendation trained with negative sampling. In: Proceedings of the 17th ACM Conference on Recommender Systems. pp. 116--128 (2023)

\bibitem{ren2019repeatnet}
Ren, P., Chen, Z., Li, J., Ren, Z., Ma, J., De~Rijke, M.: Repeatnet: A repeat aware neural recommendation machine for session-based recommendation. In: Proceedings of the AAAI conference on artificial intelligence. vol.~33, pp. 4806--4813 (2019)

\bibitem{fpmc}
Rendle, S., Freudenthaler, C., Schmidt-Thieme, L.: Factorizing personalized markov chains for next-basket recommendation. In: Proceedings of the 19th international conference on World wide web. pp. 811--820 (2010)

\bibitem{romanov2023time}
Romanov, A., Lashinin, O., Ananyeva, M., Kolesnikov, S.: Time-aware item weighting for the next basket recommendations. In: Proceedings of the 17th ACM Conference on Recommender Systems. pp. 985--992 (2023)

\bibitem{firstrepro}
Shao, Z., Wang, S., Zhang, Q., Lu, W., Li, Z., Peng, X.: A systematical evaluation for next-basket recommendation algorithms. In: 2022 IEEE 9th International Conference on Data Science and Advanced Analytics (DSAA). pp. 1--10. IEEE (2022)

\bibitem{sberpaper}
Shevchenko, V., Belousov, N., Vasilev, A., Zholobov, V., Sosedka, A., Semenova, N., Volodkevich, A., Savchenko, A., Zaytsev, A.: From variability to stability: Advancing recsys benchmarking practices. arXiv preprint arXiv:2402.09766  (2024)

\bibitem{singh2020prediction}
Singh, A., Hosein, P.: On the prediction of possibly forgotten shopping basket items. In: Artificial Intelligence and Applied Mathematics in Engineering Problems: Proceedings of the International Conference on Artificial Intelligence and Applied Mathematics in Engineering (ICAIAME 2019). pp. 559--571. Springer (2020)

\bibitem{bert4rec}
Sun, F., Liu, J., Wu, J., Pei, C., Lin, X., Ou, W., Jiang, P.: Bert4rec: Sequential recommendation with bidirectional encoder representations from transformer. In: Proceedings of the 28th ACM international conference on information and knowledge management. pp. 1441--1450 (2019)

\bibitem{vaswani2017attention}
Vaswani, A., Shazeer, N., Parmar, N., Uszkoreit, J., Jones, L., Gomez, A.N., Kaiser, {\L}., Polosukhin, I.: Attention is all you need. Advances in neural information processing systems  \textbf{30} (2017)

\bibitem{wang2022efficiently}
Wang, B.L., Schelter, S.: Efficiently maintaining next basket recommendations under additions and deletions of baskets and items. arXiv preprint arXiv:2201.13313  (2022)

\bibitem{sasrecrl}
Xin, X., Karatzoglou, A., Arapakis, I., Jose, J.M.: Supervised advantage actor-critic for recommender systems. In: Proceedings of the Fifteenth ACM International Conference on Web Search and Data Mining. pp. 1186--1196 (2022)

\bibitem{dnntsp}
Yu, L., Sun, L., Du, B., Liu, C., Xiong, H., Lv, W.: Predicting temporal sets with deep neural networks. In: Proceedings of the 26th ACM SIGKDD International Conference on Knowledge Discovery \& Data Mining. pp. 1083--1091 (2020)

\end{thebibliography}

\end{document}